\documentstyle[12pt,aasms4,epsf]{article}                   
\newcommand{\mettips}[3]
{
\noindent
\begin{figure}[t]
  \begin{center}
  \leavevmode
  \epsffile{#3}
  \end{center}
  \caption{\protect\small \it  #2 \label{#1}}
\end{figure}
}
\begin{document}

\title{Parallelization of a Code for the Simulation
of Self-gravitating Systems in Astrophysics.
Preliminary Speed-up Results.}
\author{R. Capuzzo--Dolcetta }
\affil{Istituto Astronomico, Universit\`a ``La Sapienza'', Rome, Italy.\\
E-Mail: dolcetta@astrmb.rm.astro.it}

\and

\author{P. Miocchi }
\affil{Dipartimento di Fisica, Universit\`a ``La Sapienza''  Rome, Italy.\\
E-Mail: miocchi@astrmb.rm.astro.it}

\begin{abstract}
We have preliminary results on the parallelization of a Tree-Code for
evaluating  gravitational forces in N-body astrophysical systems. For our T3D
CRAFT implementation, we have obtained an encouraging speed-up behavior, which
reaches a value of 37 with 64 processor elements (PEs). According to the
Amdahl'law, this means that about 99\% of the code is actually parallelized.
The speed-up tests regarded the evaluation of the forces among 
$N = 130,369$ particles distributed scaling the actual distribution of a
sample of galaxies seen in the Northern sky hemisphere. Parallelization of
the time integration of the trajectories, which has not yet been taken into
account, is both easier to implement and not as fundamental. 
\end{abstract}

\section{Introduction and Scientifical Motivations}
Super computers are allowing a rapid development of numerical simulations of 
large N--body systems in Astrophysics. These systems are generally composed by
both collisionless matter (such as: stars, galaxies, ...) and collisional
matter (i.e. gas).
Both phases are usually characterized by being self--gravitating, that is the
dynamics of the bodies (stars or fluid elements) is strongly influenced by the
gravitational field produced by the bodies themselves.

This {\it self-influence} is what makes the evaluation of the long--range
gravitational force the heaviest computational task to perform in a dynamical
simulation. In fact, the number of terms which has to be considered in a
direct and trivial evaluation of all the interactions between bodies grows
like $N^2$, and since many astrophysically realistic simulations require very
large $N$ (greater than $10^5$), such a direct numerical evaluation seems hard
to face with presently available computers.

To overcome this problem various approximate techniques to compute 
gravitational interactions have been proposed. Among them, the Tree--code
algorithm proposed by Barnes \& Hut\footnote{ 
Barnes J., Hut P. ``A hierachical $O(N\log N)$ force calculation algorithm''.
{\it Nature}, vol. 324, p. 446 (1986).}
is now widely used in
Astrophysics because it does not require any spatial fixed grid (like, for
example, methods based on the solution of Poisson's equation). This makes it
particularly suitable to follow very inhomogeneous and variable (in time)
situations, typical of self-gravitating systems out of equilibrium. 
In fact its intrinsic capability to give a rapid evaluation of forces allows
spending more CPU-time to follow fast dynamical evolution, in contrast to
other higher accuracy methods that are more suitable for other physical
situations, e.g. dynamics of polar fluids, where the Coulomb term is present.

With the help of the parallelization of our codes, we intend to increase by
one or two order of magnitude the number of particles we can use to represent
physical systems, in respect to that generally adopted on serial computers
($\sim 10^4$). In particular our first scientifical aim is the study of close
encounters between massive black holes and globular clusters. These latter are
systems formed by more than $10^5$ stars gravitationally bounded in a
spherical peaked distribution. Such a problem is important in the effort to
understand better the nature and formation mechanisms of the {\it Active
Galactic Nuclei}\ \footnote{
Capuzzo-Dolcetta R., Miocchi P., ``Galactic Nuclei Activity Sustained by
Globular Cluster Mass Accretion'', {\it PaSS} (1998) in press.}.
We hope parallelization makes possible to represent each
star with a single particle, in a one--to--one correspondence. This fact
clearly will make simulations much more physically meaningful.

\section{The Tree--code}
We built our own serial implementation of the Tree--code. We give here only a 
very brief description of the algorithm\footnote{
For more details see Miocchi, P. ``Costruzione di un codice numerico per lo
studio della dinamica e idrodinamica di sistemi auto--gravitanti in
Astrofisica'', Graduation Thesis, Univ. of L'Aquila (1994).}.

First, the entire system of particles (which can represent stars, single
galaxies as well as fluid elements of a gaseous self-gravitating phase) is
enclosed in a cubic box (the ``root'' box). This box is then subdivided into
eight sub-cells of half size. The subdivision continues recursively for each
of these sub-cells until one obtains cells with only one particle inside
(called terminal cells). In this way the subdivision is {\it locally} as
refined as the particle density is high. Moreover empty cell are not 
subdivided further.

Then, for each box various multipolar coefficients (total mass, position of 
the center of mass, quadrupole mass tensor, and so on) are calculated. They
will be useful to evaluate the gravitational field that the box produces on a
particle. Such coefficients, plus other useful data, are stored into pointed
arrays which are structured as a {\it tree} graph: the root box points to its
non-empty sub-cells, which point to their non-empty sub-sub-cells and so on;
terminal boxes are the ``leaves'' in this tree structure. We refer to the
above steps as the `Tree-setting' phase. Furthermore, we say 
that the root box is at the level 0 of subdivision, its sub-cells are at level
1, and so on.

In the `Force-evaluation' phase, for each particle one considers all the cells 
previously found, ``ascending'' the tree structure, starting from the root
cell, in the following way: given a cell, if it is sufficiently distant from
the particle, then the field at the particle position is evaluated by means of
a truncated multipolar expansion (using the coefficients stored for this cell
in the previous phase). Otherwise one ``opens'' the cell, passes to the next
subdivision level and considers its sub-cells. The field produced by terminal
cells are evaluated just by summing directly the contribution due to the
particle contained there.

While the direct evaluation of forces scales as $N^2$, in the Tree--code the
CPU-time requirement scales as $N\log N$, making possible simulations with high
$N$. Moreover the main advantage of this {\it gridless} method is its
adaptability to any geometrical configuration of the masses and its Lagrangian
nature. It is based on a particle representation of the density field, which
permits, for example, to calculate easily gravitational self--interaction
among the various parts of a fluid, just by representing its fluid elements
with particles as well. This gives the possibility of using Lagrangian 
methods such as the SPH\footnote{ 
Monaghan J.J., {\it ARA\&A}, vol. 30, p. 543 (1992)}
for simulating hydrodynamical
evolution. A disadvantage is the needing to re-execute all the phases
previously described at every time-step during the simulation, even if the
particles had small displacements.

Recently a new method has been proposed; it is called Fast Multipole Algorithm
\footnote{
Greengard L., ``The Rapid Evaluation of Potential Field in Particle Systems'',
PhD Thesis, MIT Press (Cambridge, MA, London, UK) (1987).}
(FMA). Its CPU-time is claimed to scale as $N$, at least in
quasi-homogeneous 2-D particle distributions. Were this linear behavior
confirmed in 3-D highly non-uniform cases, the FMA would really be appealing
for use in astrophysical simulations. For this reason we compared CPU times of
our own serial implementations of an adaptive 3-D FMA and a Tree--code to
evaluate gravitational forces among $N$ particles in 
several (uniform and clumped) spatial configurations. These comparison tests
(see [\cite{1}] and [\cite{2}]) indicates the Tree--code as faster than the FMA in
all the situations considered for $N$ up to $2\cdot 10^5$. This convinced us
to concentrate our efforts in the parallelization of the Tree--code.

\section{The Parallelization}
Tree--code is difficult to parallelize mainly because the evaluation of all
the interactions among bodies is not completely separable into a set of
independent tasks. 
The difficulties are due mainly to the following peculiarities:
\begin{enumerate}
\item \label{one} gravitation is a long--range interaction: inter-processor
communications are inevitable;
\item \label{two} non-uniform distributions (very frequent in Astrophysics)
mean great  differences in the amount of contributions to the force on each
particle: a good load balancing is hard to be achieved.
\end{enumerate}
Point \ref{one} means that we should perform an appropriate domain
decomposition among PEs in order to minimize inter-communications. This is
generally obtained with a domain subdivision which assigns to PEs domains
which are as spatially contiguous as possible. This consideration is
particularly important in a message passing context. 
We will face with this kind of parallelization in the future. At present we
want to exploit the great transfer rate of the T3D/E machines, parallelizing
our serial code on the T3D using CRAFT language without caring too much of
reaching an optimal data locality. Anyway we must pay attention to point
\ref{two} which, in a message passing approach, implies that the domain
decomposition should be ``weighte'' in order to take into account the work
load for each PE.

We found that the greatest difficulties in getting good performances are in
the Tree-setting phase, in which is not easy to avoid load unbalancing and
overhead times due mainly to barriers and critical regions. 
In order to eliminate such synchronization ``bottlenecks'' we adopted the 
following scheme:
\begin{enumerate}
\item All PEs work together to build the tree data and pointers structure
starting
from the root and up to a certain level of subdivision, say $L$. This is the
lowest level such that the number of non-empty cells found (they have size
$l/2L$, being $l$ the root cell size) is greater than $kp$, with $p$ the
number of PEs and $k > 1$ a coefficient. In this phase we have reached a well
balanced work load, exploiting the fact that the maximum number of non-empty
cells at the level $L$, about $8^L$, is not too large from the point of view of
the memory occupation. This cannot be done for the complete tree structure 
because, in general, astrophysical distributions are very clumped and one can easily 
reach more than 10 levels of subdivisions, which means to have, at least in
principle, more than $8^{10}$ possible cells with their corresponding memory
locations!
We won't give here further details on this argument, anyway the interested
reader can find some explications in [\cite{3}].
\item At this point, a work distribution is performed by means of do-shared loops 
distributed on the cells of level $L$. In this way each PE works
{\it independently} on its 
cells and it builds, for them and all their {\it descendants}, the related
tree structure. In this phase, the work load for a PE exploring a cell, is
determined by the total number of non-empty descendants it finds. Thus, load
balancing is guaranteed by fixing $k$ high enough such that every PE has a
total work load which is constant and equal to all other PEs, apart from
statistical fluctuations (see next section).
\end{enumerate}

In order to make a good distribution not only of the work but also of the
data, the vectors and the various arrays which reproduce the tree data
structure into the memory, have to be shared. This means that, since each PE
which is working on its cells has to update frequently such shared arrays, in
order to avoid race conditions one should use {\it atomic updating} or even
critical regions. This would give a very bad speed-up, as we verified. Thus
we prefer to use an alternative approach which we can call ``double passing''.

In such a scheme, in the first passage all PEs explore their cells as they
should stored and updating arrays, but without doing it. They just find how
many locations they use in these arrays. Then they subdivide arrays and
vectors into segments, such that each PE will use exclusively its own segment.
Obviously, the PEs have to communicate one each other the respective
boundaries of the array segments used, but this is done only once. Then, in
the second passage, each PE repeats the scanning of the cells storing proper
data and pointers only on its own memory locations. This permits to avoid any
atomic updating of shared pointers, race conditions, critical regions, and
so on.

\section{Results and Future Prospects}
To test the speed-up of our parallelized code, we distributed $N = 130,329$
particles scaling the density distribution of a sample of galaxies in the
Northern galactic hemisphere (see Fig.\ref{fig1}), taken from the Leda
catalog\footnote{
Di Nella M., Paturel G., {\it Comptes Rendue de l'Acad. des Sciences de Paris},
Sec. II, t319, p. 57 (Paris, FR, 1994).}. As one can see, this is a very
clumped distribution which constitutes a good benchmark for a test of the code. 

\mettips{fig1}{Observed distribution of a sample of galaxies in the
sky Northern hemisphere (Leda catalog). The axes units correspond to about 80
Mpc.}{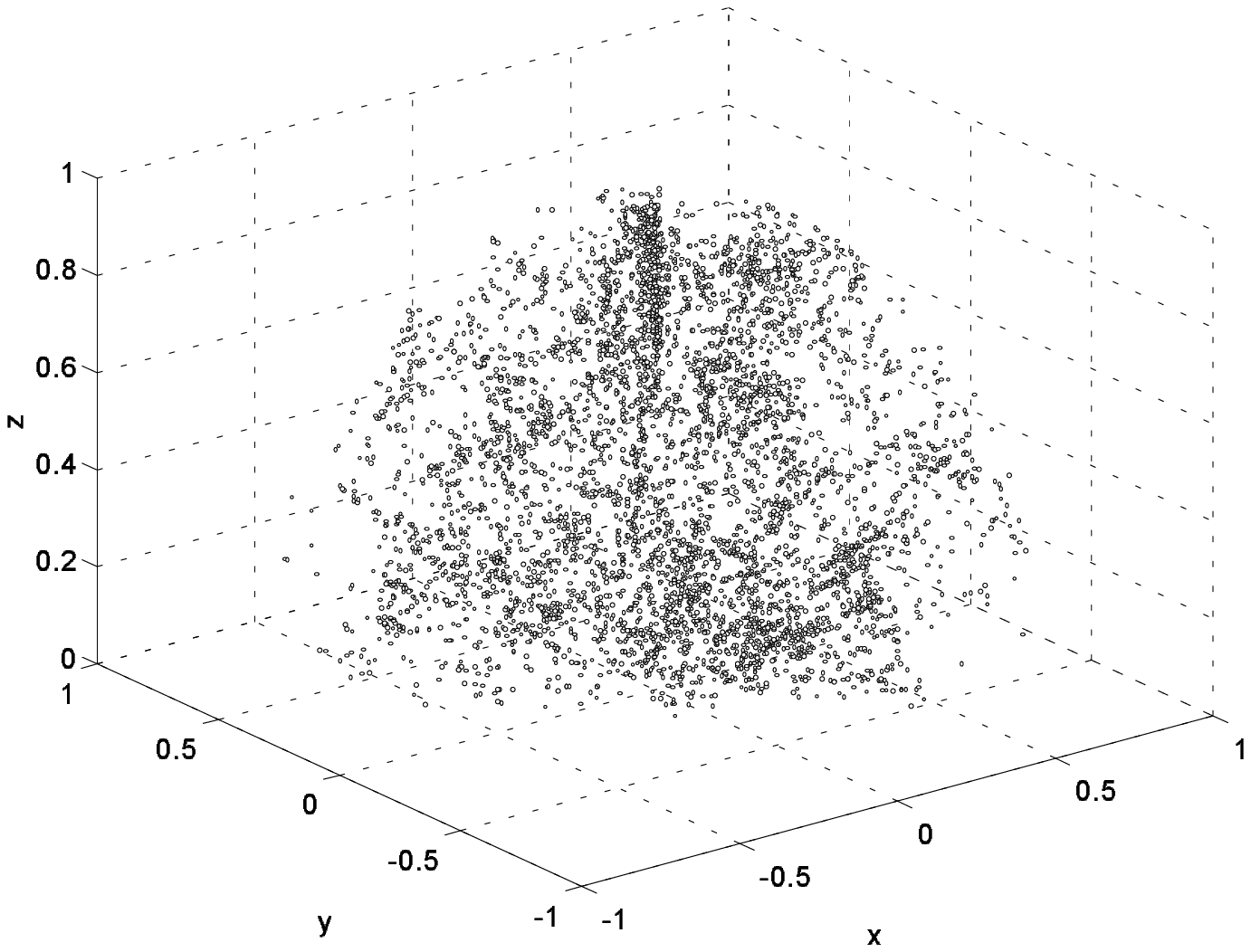}

Moreover, we fixed $k = 30$ and the `tolerance' parameter $a = 0.7$ in the
force calculation. This latter parameter regards the criterion to establish
whether or not in the Force-evaluation phase, a cell is `sufficiently
distant' from a particle, as we have described.

In Fig.\ref{fig2} we show the speed-up results obtained on the T3D. We give
the total speed-up, that regarding the Tree-setting and that regarding the
Force-evaluation. The curves in Fig.\ref{fig2} confirm the Tree-setting as the
most difficult part of the algorithm to be well parallelized, while the
Force-evaluation speed-up has a very good behavior in spite of the fact it
uses intensively remote reading. In fact, this latter phase has been trivially
parallelized distributing work among PEs, by means of a do-shared loop on 
the particles. Each PE calls its private subroutine to evaluate the force
acting on its own particles. To do this it has to read the data of the various
cells which are stored, in general, in the memory of another PE.

\mettips{fig2}{Measured speed-up for the Tree-setting phase (blue),
the Force-evaluation phase (green) and total (red). The test refers to the
evaluation of gravitational forces among N = 130,139 particles (see text).}
{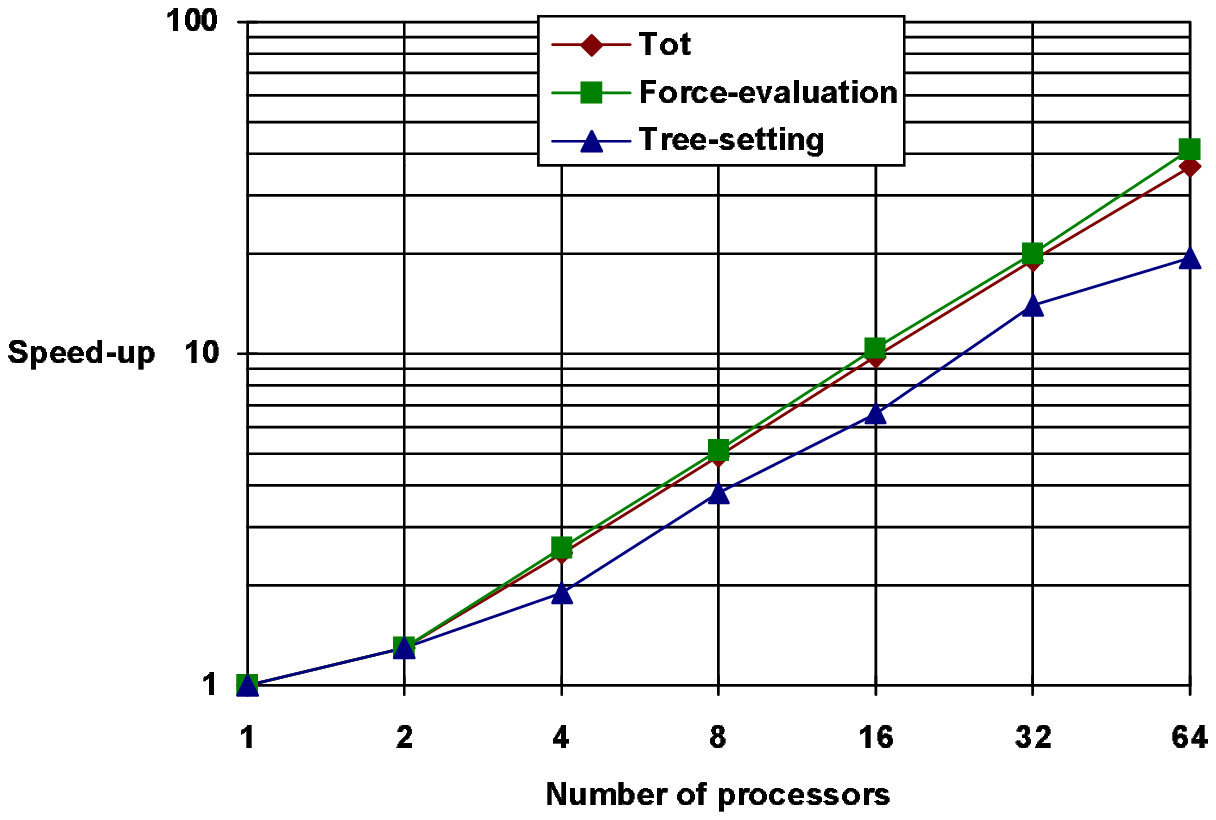}

In Fig.\ref{fig3} the work load distribution is shown for the run with 8
processors, during both phases. The work load has been normalized to the
average over all PEs. For the Tree-setting phase it has been calculated as
proportional to the total number of non-empty descendants each PE finds
starting from its own cells at the level L. In the Force-evaluation phase it
has been deduced from the total number of force contributions each PE has to
sum in considering all the particles in its domain. Note how load balancing 
is very good for this latter phase, while for the Tree-setting there are
differences among PEs work load which, in any case, do not exceed the 20\% of
the average.

\mettips{fig3}{
Normalized work load distribution over 8 processors in both phases: Tree- 
setting (red) and Force-evaluation (green).}
{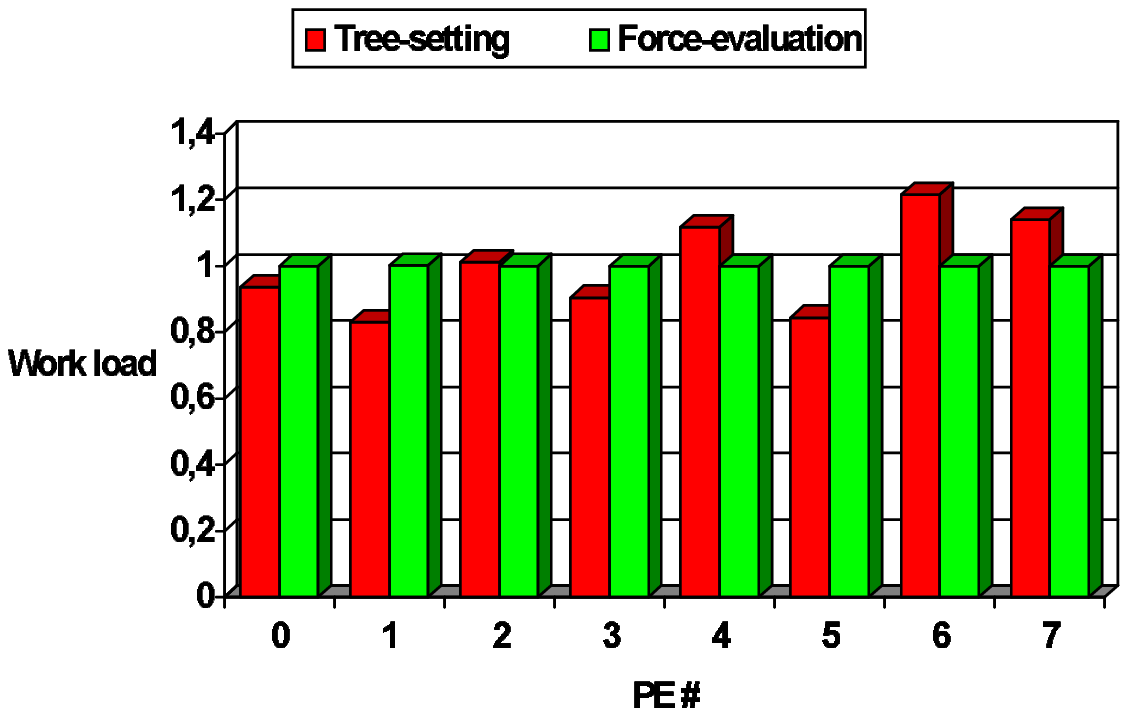}

To conclude, the results are rather good: the total speed-up is high enough
and it does not show any flattening, at least for $p\leq 64$. According to the
Amdahl's law, this indicates that the effective parallelized portion of the
code (whose CPU-time scales like $1/p$) is about 99\% of the total. One has
to consider also that for $p > 16$ the amount of particles per processor is
not that high (less than 5,000). We think that using more particles we would
get even better results. This drives us to extend the parallelization 
also to the time integration of particles trajectories and to the SPH routines
for the numerical simulation of hydrodynamics of a self--gravitating fluid.
In any case these latter goals are quite easier to be reached than that of the proper 
parallelization of the Tree--code we have done here.
\section{Acknowledgments}
This work has been supported by the grant {\it k90rmzz1} at the CINECA
Supercomputing center (Bologna, Italy). We thank warmly dr. M. Voli (CINECA)
for his valuable help and suggestions and dr. M. Montuori (Univ.
``La Sapienza'') for giving us the data sample of galaxies positions.

\end{document}